\documentclass[a4paper]{cas-sc}

\usepackage[authoryear,longnamesfirst]{natbib}

\usepackage{amsmath}
\usepackage{hyperref}
\usepackage{graphicx}
\usepackage{subfigure}
\usepackage{algorithm}
\usepackage{algorithmic}
\usepackage{booktabs}
\usepackage{threeparttable}
\usepackage{multirow}
\usepackage[figuresright]{rotating}
\usepackage{overpic} 
\usepackage{array}

\def\tsc#1{\csdef{#1}{\textsc{\lowercase{#1}}\xspace}}
\tsc{WGM}
\tsc{QE}
\tsc{EP}
\tsc{PMS}
\tsc{BEC}
\tsc{DE}

\ExplSyntaxOn
\cs_gset:Npn \__first_footerline:
  { \group_begin: \small \sffamily \__short_authors: \group_end: }
\ExplSyntaxOff 

\begin{document}
\let\WriteBookmarks\relax
\def\floatpagepagefraction{1}
\def\textpagefraction{.001}

\shorttitle{Towards computing complete parameter ranges in parametric modeling}


\title [mode = title]{Towards computing complete parameter ranges in parametric modeling}                      



%
\author[1]{Zhihong Tang}[style=chinese]


\author[1]{Qiang Zou}[style=chinese]

\author[1]{Shuming Gao}[style=chinese]

\cormark[1]

\fnmark[3]

\ead{smgao@cad.zju.edu.cn}


\credit{Conceptualization of this study, Methodology, Software}

\credit{Data curation, Writing - Original draft preparation}

\cortext[cor1]{Corresponding author}
\cortext[cor2]{Principal corresponding author}



\begin{abstract}
In parametric design, the geometric model is edited by changing relevant parameters in the parametric model, which is commonly done sequentially on multiple parameters. Without guidance on allowable parameter ranges that can guarantee the solvability of the geometric constraint system, the user could assign improper parameter values to the model’s parameters, which would further lead to a failure in model updating. However, current commercial CAD systems provide little support for the proper parameter assignments. Although the existing methods can compute allowable ranges for individual parameters, they face difficulties in handling multi-parameter situations. In particular, these methods could miss some feasible parameter values and provide incomplete allowable parameter ranges. To solve this problem, an automatic approach is proposed in this paper to compute complete parameter ranges in multi-parameter editing. In the approach, a set of variable parameters are first selected to be sequentially edited by the user; before each editing operation, the one-dimensional ranges of the variable parameters are presented as guidance. To compute the one-dimensional ranges, each variable parameter is expressed as an equality-constrained function, and its one-dimensional allowable range is obtained by calculating the function range. To effectively obtain the function range which can hardly be calculated in a normal way, the function range problem is converted into a constrained optimization problem, and is then solved by Lagrange multiplier method and the Niching particle swarm optimization algorithm (the NichePSO). The effectiveness and efficiency of the proposed approach is verified by several experimental results.
\end{abstract}




\maketitle

\section{Introduction}

Parametric modeling has been the most widely applied CAD paradigm. In parametric CAD systems, an object is represented by a parametric model and a geometric model consistent with each other \citep{Shah1995parametric}. The geometric model is edited and updated by changing the values of relevant parameters sequentially in the parametric model. In real-world applications, the model update sometimes suffers failure. An indispensable cause of such failure lies in improper assignments of parameter values, which results in an unsolvable geometric constraint system. To avoid this, guidance of allowable parameter ranges is needed to guarantee the solvability of a geometric constraint system. 

Parameter range computation for solvability is a long-standing problem of great importance in parametric modeling. As a geometric constraint system can be represented as an algebraic equation system, computing parameter ranges is essentially projecting the high-dimensional semi-algebraic set defined by the equation system to the low-dimensional one related to parameters. This problem can be handled by cylindrical algebraic decomposition method \citep{tarski1998decision,collins1975quantifier,arnon1984cylindrical}, but current algorithms have difficulties in dealing with practical CAD problems due to the high time complexity. Thus, researchers are seeking a more efficient way to calculate parameter ranges. Existing methods \citep{joan2001applying,sitharam2011cayley1,sitharam2011cayley2,sitharam2014beast,van2005constructive, hidalgo2012computing,hidalgo2014reachability,hidalgo2015hgraph} can compute allowable ranges of individual parameters in specific geometric constraint systems. However, when applied to the multi-parameter editing, the methods could omit certain feasible parameter values, which will result in incomplete parameter ranges. 

To compute complete parameter ranges in multi-parameter editing of 2D geometric constraint systems, a systematic approach is proposed in this paper. Considered 2D geometric constraint systems are formed by dimensional constraints (e.g. distance and angle), structural constraints (e.g. perpendicular and parallel), and algebraic constraints that relate the parameters from different parametric constraints (e.g. $d_1=2d_2$ where $d_1$ and $d_2$ are two distance parameters); integral constraints that involve integral symbol in the equations are not considered in the present work. The parameters whose ranges are to be computed are defined by dimensional constraints. In the approach, parameters to be edited are first selected by the user as variable parameters, and are then edited in a sequential manner; before each editing operation, the one-dimensional allowable range of each variable parameter is provided to the user as guidance. To compute one-dimensional allowable ranges, each variable parameter is expressed as an equality-constrained function; the function range, which is equivalent to the one-dimensional parameter range, is then obtained by computing endpoint candidates and determining valid intervals. The completeness of the computed parameter ranges is guaranteed by maintaining a complete parameter space, which is achieved by involving all the geometric constraints as constraint equations except for those related to unassigned parameters. By sequentially editing the variable parameters within the computed one-dimensional allowable ranges, the solvability of the geometric constraint system can be guaranteed. The main features of the proposed approach include: (1) the approach can provide complete one-dimensional ranges of variable parameters that underwent editing in the multi-parameter editing problems; and (2) the approach is not restricted to specific geometric constraint systems.

The rest of this paper is organized as follows. Section 2 reviews related work on geometric constraint solving and parameter range computation. Section 3 presents the overview of the proposed approach. Section 4-6 describe the critical steps of the proposed approach, the algebraic representation construction of variable parameter, the computation of closed endpoint candidates, and the computation of open endpoint candidates, respectively. Section 7 provides several case studies to demonstrate the effectiveness and efficiency of the proposed approach. Finally, Section 8 concludes the paper and presents future work.

\section{Related work}
\subsection{Geometric constraint solving}

To compute parameter ranges that guarantee the solvability of a geometric constraint system, the main difficulty lies in constructing the relation between the solvability of the system and the values of the parameters. Generally, a geometric constraint system is considered solvable if it can be solved by one of the geometric constraint solving methods. Hence, basic concepts, methods and techniques of geometric constraint solving are recalled. Basically, a geometric constraint system, embedded in a 2D or 3D space, is formed by a set of geometric entities and a set of geometric constraints; the main goal of geometric constraint solving is to find the coordinates of these geometric entities that satisfy all the geometric constraints.

A significant number of studies on geometric constraint solving have been reported \citep{bettig2011geometric}. In this paper, the geometric constraint solving methods are classified into two categories, equation-based methods and constructive methods.

The equation-based methods basically consider a geometric constraint system as a system of nonlinear algebraic equations, and aim at finding its roots. The equation-based solving methods can be further divided into two subgroups: numerical methods and symbolic methods. Numerical methods solve an equation system by iterations, and obtain one or more sets of numerical values as the solutions \citep{Sommese2005TheNumerical}. One set of methods focus on finding a root quickly, e.g. Newton-Raphson iteration \citep{light1982modification, lin1981variational,nelson1985juno} and relaxation method \citep{sutherland1964sketchpad,Hillyard1978Characterizing,borning1981programming}. The other set of methods aim at finding all roots, e.g. homotopy continuation method \citep{Allgower1993homotopy,lamure1996homotopy,Durand1998Symbolic,imbach2017homotopy} and subdivision methods \citep{DOKKEN1985Finding,Sherbrooke1993Computation,elber2001geometric,BARTON2011solving}. There is always a trade-off between efficiency and exhaustive searching. Numerical methods are general since they work at the equation level, but they are time-consuming without involving constraint decomposition, especially for large-scale systems. Symbolic methods employ symbolic techniques to convert an equation system into a triangular form through, for example, Gr\"{o}bner bases \citep{bose1995grobner,buchanan1993constraint, kondo1992algebraic} and Wu-Ritt decomposition \citep{chou1988introduction, chou1988mechanical}. Symbolic methods are capable of finding all roots, but are rarely employed in practical CAD problems due to their high computational load.

The constructive methods aim at finding a constructive plan for the geometric constraint system so as to rebuild the geometric model in a step-by-step manner. The constructive plan decomposes the whole geometric constraint system into a set of subproblems that can be solved independently, and gives a sequence to recombine them together. The constructive methods include the rule-based methods and the graph-based methods. The rule-based methods use a group of rules to aggregate subsystems into bigger ones by which the construction steps are readily available \citep{dufourd1998geometric, joan1997correct}. However, it is hard, if not impossible, to make the rules exhaustive \citep{jermann2006decomposition}. Another important category of geometric constraint system decomposition is graph-based, which abstracts a geometric constraint system into a graph, and then uses existing graph algorithms to find the constructive patterns in the graph (and therefore the geometric constraint system). The graph-based methods are pioneered by Serrano \citep{serrano1987constraint}, and have been much developed in a series of research studies \citep{barford1987graphical, owen1991algebraic, fudos1994Correctness, bouma1995geometric, fudos1997graph, gao2004solving}. Moreover, a few studies \citep{Hoffman1995geometric,Gao2006ctree,van2010anonrigid} focused on extending the existing research results of geometric constraint solving from 2D systems to 3D ones, which still remains an open issue due to the lack of available rigidity theory in 3D.

Regarding the two categories of geometric constraint solving methods, the conditions of keeping a system solvable can be defined as follows. 

For the constructive methods, a geometric constraint system is solvable only if all subproblems in the constructive plan can be solved; otherwise, it fails to solve the system. On this basis, the parameter ranges to be computed are the values where all the subproblems keep solvable. The difficulty lies in listing the solvable conditions for all types of subproblem patterns. Van der Meiden and Bronsvoort \citep{van2005constructive} discussed two subproblem patterns, triangles in 2D and the tetrahedrals in 3D. Nevertheless, the subproblem patterns are far from complete.

Regarding the equation-based methods, a geometric constraint system is regarded as solvable if its equivalent equation system has at least one root; otherwise, it is regarded unsolvable. The cause of the unsolvable system lies in that the multivariate implicit hyper-surfaces defined by the constraint equations have no intersection. Thus, the parameter ranges to be computed need to be the values where these hyper-surfaces maintain intersected. 

\subsection{Parameter range computation}

When users need guidance on parameter assignments to retain the nature of a geometric constraint system or a CAD model, the parameter range computation is used. Thus far, the problem of parameter range computation has two categories: problem for valid models and problem for solvable geometric constraint systems.

\subsubsection{Problem for valid models}

A valid model is a model that has the same topology as a given prototype model \citep{van2010tracking}. When editing parameters, in order to mostly retain the features and functions of the model, users need guidance on parameter assignments to maintain the model topology.

This problem arose from discussions about the definition of a parametric family of solids \citep{shapiro1995parametric}. Shapiro and Raghothama \citep{raghothama1998boundary} provided a formal definition of a parametric family of solids by the continuity principle: small changes in parameter values result in small changes in the b-rep. This definition provides users with slight guidance on parameter editing, but cannot completely prevent users from assigning inappropriate parameter values that results in topology change. 

Later, it has been considered to provide users with valid parameter ranges, which guarantees to regenerate a valid model in the same parametric family as the former one.

A pioneer work has been proposed by Hoffman and Kim \citep{hoffmann2001towards}. Simple 2D geometric constraint systems, consisting of horizontal and vertical line segments with distance constraints between parallel line segments, were considered. One-dimensional valid ranges of distance parameters are determined for keeping the topology of line segments unchanged. Jiang et al. \citep{Jiang2003valid} went a step further on the basis of the work by Hoffman and Kim. Considering the same type of geometric constraint systems in \citep{hoffmann2001towards}, they proposed an algebraic algorithm to determine the valid parameter space in multi-parameter editing. 

Van der Meiden and Bronsvoort \citep{van2010tracking} proposed a method to compute valid ranges of individual parameters in 3D parametric and feature-based modeling. The basic idea of the method is building a 3D geometric constraint system relating parameters and topology entities to construct degenerate cases where the model topology may change. The method relies on cellular models \cite{Bronsvoort1998Representation}.

\subsubsection{Problem for solvable models}

Parameter range computation for solvable geometric constraint systems aims at determining parameter ranges that keeps the system solvable. This problem is algebraically equivalent to calculating the allowable ranges of parameters where the equation system has at least one root. In single-parameter editing, the parameter range is a set of disjoint intervals; while in multi-parameter editing, the parameter ranges are essentially a subset of high-dimensional space with complex boundaries. The parameter ranges can be regarded as the projection from the semi-algebraic set defined the constraint equations, thus can be generally handled by the cylindrical algebraic decomposition method \citep{tarski1998decision,collins1975quantifier,arnon1984cylindrical}. However, as the method has doubly-exponential time complexity \citep{collins1975quantifier}, it is not applicable to practical problems in parametric modeling.

Joan-Arinyo and Mata \cite{joan2001applying} presented a method to compute parameter ranges under the condition that intervals are assigned to parameter values. Geometric constraint systems that are ruler-and-compass constructible are considered. The method is based on numerical sampling.

Meera Sitharam et al. \citep{sitharam2011cayley1,sitharam2011cayley2,sitharam2014beast} concerned realization space of 1-dof 2D linkages, which can be abstracted as a 1-dof geometric constraint system with only points and distance parameters between them; algorithms were given to obtain distance intervals between two points for generic 1-dof tree-decomposable linkages.

Van der Meiden and Bronsvoort \citep{van2005constructive} proposed a constructive method to compute allowable parameter ranges. The constructive method consists of two phases. First, by constructing degenerate cases for all subproblems, the critical values of the variable parameter where subproblem solvability changes are determined. Then, the critical values constitute a group of interval candidates, and the validity of each interval is evaluated by testing the system solvability at a sampling point within the interval. Regarding single-parameter editing problem, the constructive method has good performances, but the geometric constraint systems considered are those that can be decomposed into 2D triangular or 3D tetrahedral subproblems. 

Hidalgo and Joan-Arinyo \citep{hidalgo2012computing,hidalgo2014reachability} formalized the constructive method \citep{van2005constructive}, proved its correctness in single-parameter editing problem, and presented an implementation of this method. Considering that the constructive method \citep{van2005constructive} relies heavily on searching the construction steps related to the variable parameter, Hidalgo and Joan-Arinyo \citep{hidalgo2015hgraph} further introduced h-graphs as a  representation for graph tree decomposition to improve the efficiency for capturing the dependencies between the construction steps of a geometric constraint problem.

In short, existing methods have the following limitations. First, only a few specific types of geometric constraint systems can be handled; and second, these methods face difficulties in multi-parameter editing.

\section{Approach overview}

In order to effectively support multi-parameter editing of a 2D geometric constraint system, a systematic approach is proposed. Considered geometric constraint systems are those formed by dimensional constraints, structural constraints, and algebraic constraints. Basically, one-dimensional intervals are adopted as the presentation scheme for allowable parameter ranges; the variable parameters to be edited are first selected by the user, and then sequentially edited within the allowable parameter ranges; also, a method is proposed to compute complete one-dimensional ranges of variable parameters.

\begin{figure*}[t]
	\centering
    \includegraphics[scale=0.55]{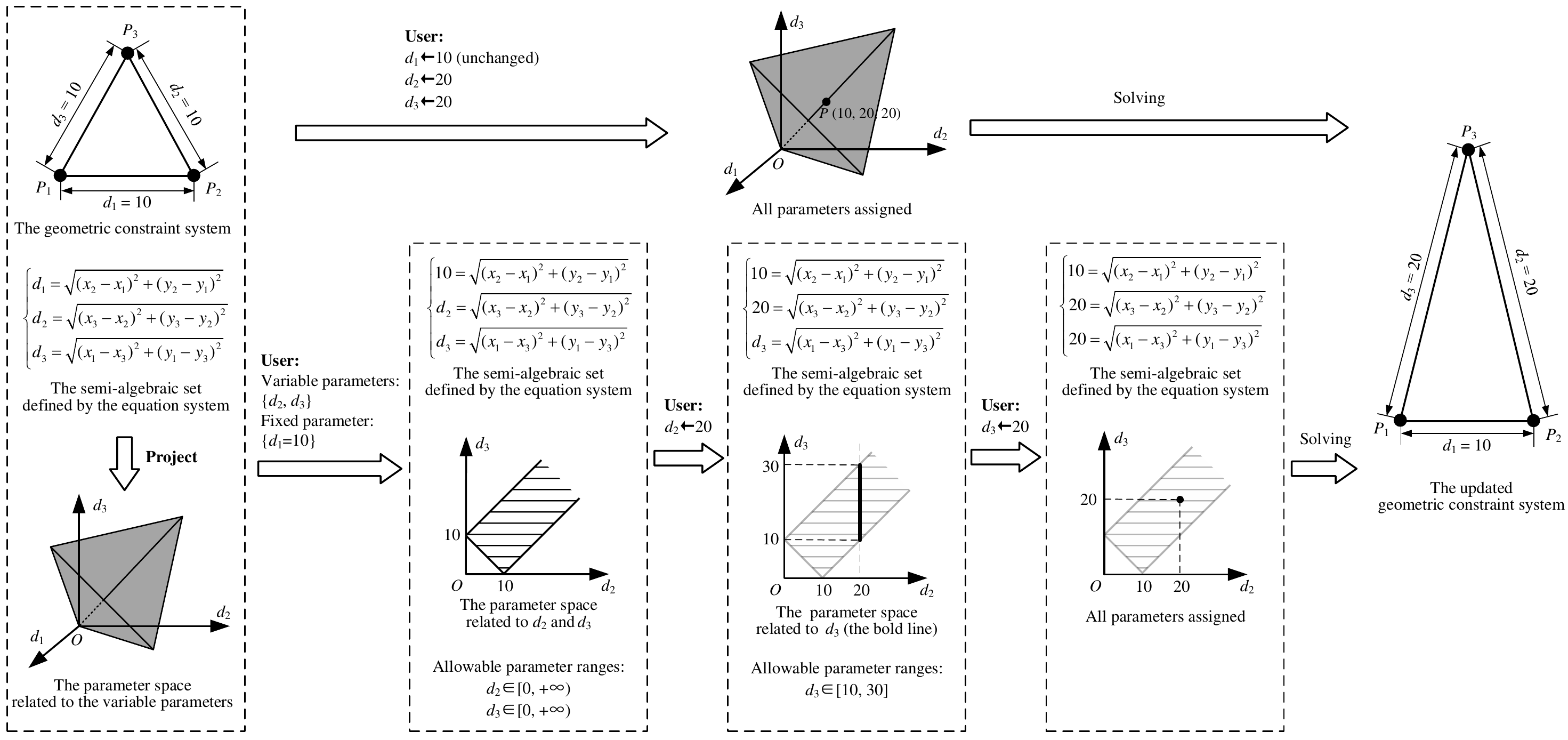} 
	\caption{Two ways of multi-parameter editing and their corresponding presentation schemes of allowable parameter ranges.}
\label{fig:3-1}
\end{figure*}

\subsection{Presentation scheme of parameter ranges}

To provide the user with the guidance of editing parameters, an appropriate presentation scheme for the allowable parameter ranges is required. Basically, there are two requirements: (1) it should conform to users' editing manner; and (2) it should be easy-to-understand for the user.

The parameter ranges essentially make up a high-dimensional space with complex boundaries. A naïve way to guide the user is to present this set directly to him/her. The user needs to set a point within the high-dimensional parameter space so that the system's solvability can be guaranteed. An example is presented in the upper pipeline of Figure \ref{fig:3-1} to illustrate this process. The geometric constraint system defining a triangle consists of three points {$P_1(x_1,y_1)$, $P_2(x_2,y_2)$, $P_3(x_3,y_3)$} and three non-negative parameters {$d_1$, $d_2$, $d_3$} defining the distances between {$P_1P_2$, $P_2P_3$, $P_1P_3$}, respectively. Let $P=(d_1,d_2,d_3)$. According to the triangle inequalities, the space of the three parameters, denoted as $S_{P}$, is expressed as follows and is plotted in Figure \ref{fig:3-1}:

	\begin{align}
    \text{$S_{P}=\{P:$}\left\{
             \begin{array}{ll}
             d_1+d_2\geq d_3\\
             d_2+d_3\geq d_1\\
             d_3+d_1\geq d_2\\
             d_1\geq 0\\
             d_2\geq 0\\
             d_3\geq 0
             \end{array}
    \right.\text{\}.}
	\label{eq:3-1}
	\end{align}
Note that greater-than-equal is used rather than greater-than, because the degenerate cases where a triangle degenerates to a line are also permitted. The system is guaranteed to be solvable if the parameter values are assigned within $S_{P}$, e.g. assigning {$d_1=10$, $d_2=20$, and $d_3=20$}.

Calculating the space of parameters, however, is not a trivial task. It is essentially the projection from the higher-dimensional semi-algebraic set defined by the equation system, but this projection is hard to build. Moreover, understanding such a high-dimensional parameter space requires the user to have a good mathematical expertise. Therefore, presenting parameter ranges in a direct scheme (as shown in the upper pipeline of Figure \ref{fig:3-1}) can hardly be the best choice for multi-parameter editing.

\begin{figure}[t]
	\centering
	\includegraphics{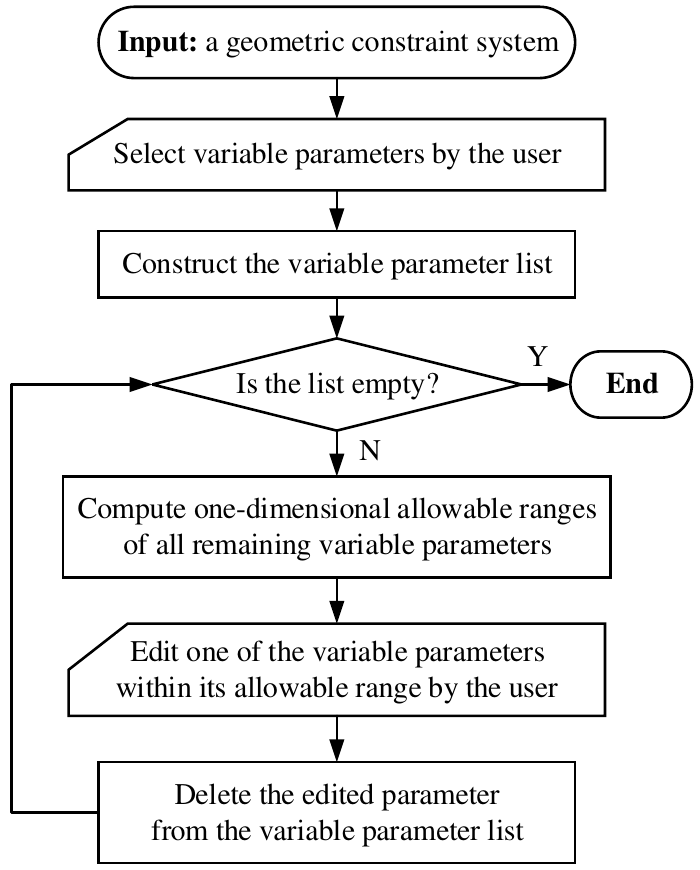}
	\caption{Flowchart of the multi-parameter editing process in 2D geometric constraint systems.}
\label{fig:3-2}
\end{figure}

Considering that users typically edit multiple parameters in a sequential manner and understand one-dimensional parameter range well, one-dimensional intervals are adopted in this paper as the presentation scheme for allowable parameter ranges. The process of multi-parameter editing of a 2D geometric constraint system with one-dimensional parameter ranges is illustrated in Figure \ref{fig:3-2}. It allows the user to edit parameters one-by-one with easy-to-understand guidance. First, a set of variable parameters, which are the parameters to be edited in this process, are selected by the user. Then, the one-dimensional allowable range of each variable parameter is automatically computed and presented to the user. The next, the user edits one of the variable parameters within its computed allowable range. After editing, the edited parameter is deleted from the variable parameter list, and the allowable ranges of the remaining variable parameters are re-computed and updated accordingly. The editing step is repeated until the list of variable parameters becomes empty. 

The above process can be regarded as creating a point by setting its coordinates one by one within the space related to variable parameters. An example is illustrated in the lower pipeline of Figure \ref{fig:3-1}. To start with, $d_2$ and $d_3$ are selected by the user as the variable parameters while $d_1$'s value stays fixed. At this first stage, the parameter space related to $d_2$ and $d_3$, plotted as a 2D region for reference, is the restriction from $S_P$ into the slice $d_1=10$; the one-dimensional allowable ranges of $d_2$ and $d_3$ should both be $[0,+\infty)$, since each of them is the projection from the 2D region to the corresponding axis. Then, suppose the user selects to edit $d_2$ as $20$; $d_2$ is removed from the list of variable parameters and $d_3$ becomes the only remaining variable parameter. At this second stage, the parameter space related to the remaining variable parameter(s) is the restriction of the former one into the slice $d_2=20$, which turns to be an one-dimensional set; thus, the allowable range of $d_3$ can be directly determined and updated as $[10,30]$. Finally, suppose that the user continues to assign $20$ to $d_3$. Since all the variable parameters have been edited within the provided allowable ranges, no parameter range needs to be provided at this stage, and the geometric constraint system is thus guaranteed to be solvable.

Compared with the previous direct scheme, the one-dimensional interval scheme is much easier to understand by the user. Thus, it is adopted as the presentation scheme of parameter ranges in our approach. The critical issue then turns to be how to compute one-dimensional ranges of the variable parameters during the multi-parameter editing process.

\subsection{Computation of complete parameter ranges}

\begin{figure}[t]
	\centering
    \subfigure[]{\includegraphics[scale=0.8]{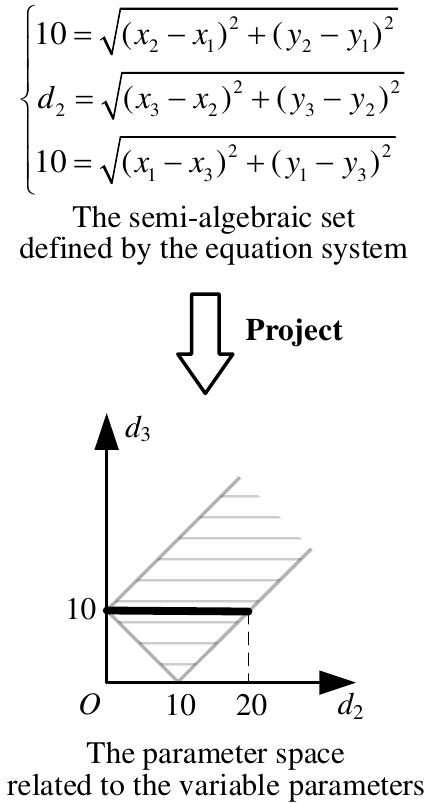}}
    \subfigure[]{\includegraphics[scale=0.8]{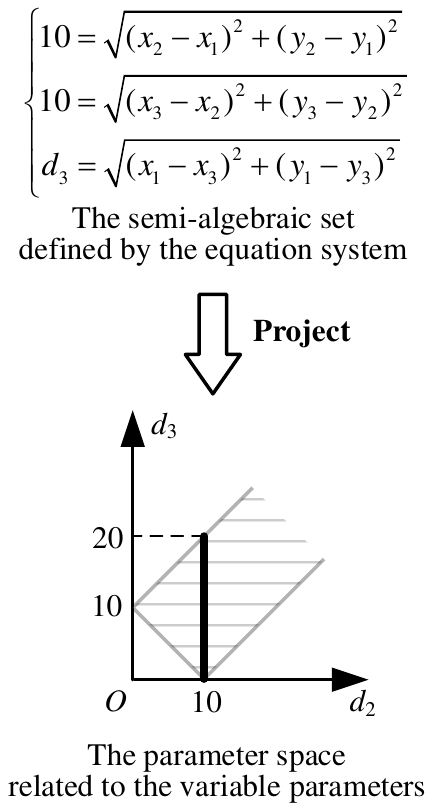}}
	\caption{Illustrations of incomplete parameter space: (a) computing range for $d_2$; and (b) computing range for $d_3$.}
\label{fig:3-3}
\end{figure}

For computing one-dimensional parameter ranges, there already exist several approaches \citep{van2005constructive, hidalgo2012computing}. These existing approaches, though mainly applied in single-parameter editing previously, can also be applied in solving multi-parameter problems. However, they turn out to output incomplete parameter ranges that omit certain feasible values. This is essentially because the computed one-dimensional ranges are projected from an incomplete parameter space. This incomplete space is resulted from unnecessarily considering the values of the unassigned variable parameters as constraints; the ideal parameter space is thus restricted into the slice defined by these constraints, thereby breaking its completeness. For better understanding, we recall the first stage of multi-parameter editing in Figure \ref{fig:3-1}, i.e. $d_2$ and $d_3$ are variable parameters and $d_1$ is the fixed parameter. Note that the transparent region is the ideal parameter space related to $d_2$ and $d_3$ for reference. Using the existing approaches \citep{van2005constructive, hidalgo2012computing} to compute parameter ranges, the range of $d_2$ turns to be $[0,20]$, since the parameter space it projected from is the restriction of the ideal one into $d_3=10$, shown in Figure \ref{fig:3-3}(a); and the range of $d_3$ also turns to be $[0,20]$ because of the restricted parameter space shown in Figure \ref{fig:3-3}(b). The computed ranges are thus shorter than the ideal ones, which could provide wrong guidance to the users.

Therefore, it is of great importance to compute complete parameter ranges. Here, the word "complete" means that the computed parameter ranges are consistent with the ideal parameter space, and does not omit any feasible parameter values. To achieve this, the critical issue to be addressed is to maintain a complete parameter space from which a parameter range is projected.

A systematic method is thus proposed in this paper to compute the one-dimensional allowable ranges of variable parameters. In this method, each variable parameter is first expressed as an equality-constrained function as follows:
	\begin{align}
	\begin{array}{c}
	p=f(X),\\
	\text{ s.t. }  G(X)=0,
	\end{array}
	\label{eq:3-2}
	\end{align}
where $p$ denotes a variable parameter, and $X$ denotes the coordinates of the geometric objects. Then, the one-dimensional allowable range of the variable parameter is obtained by calculating the function range. The completeness of the computed parameter ranges is guaranteed by maintaining a complete parameter space for projection, which is specifically achieved by involving all the geometric constraints as constraint equations except for those related to unassigned variable parameters.

The range of the equality-constrained function is a set of disjoint intervals. To determine the function range, the endpoint candidates of the intervals are calculated, including the closed endpoint candidates and the open endpoint candidates. With the endpoint candidates obtained, the valid intervals of the function range are further determined by the sampling-based method proposed by van der Meiden and Bronsvoort \citep{van2005constructive}. The sampling-based method first generates a set of interval candidates composed by two subsequent endpoint candidates, and then determines the validity of the interval candidates by examining the solvability of the geometric constraint system at a sample point within each interval candidate as well as its endpoints. Finally, by combining all the valid intervals, the one-dimensional parameter range is obtained.

In view that the key part of the proposed approach is the computation of the complete allowable range of a variable parameter, this part is described in more detail below, which consists of the following three main steps: 

\begin{enumerate}

\item
Represent the variable parameter as an equality-constrained function.

\item
Compute closed endpoint candidates of the function range.

\item
Compute open endpoint candidates of the function range.

\end{enumerate}

\section{Algebraic representation construction of a variable parameter}

The equality-constrained function of a variable parameter is symbolically translated from the equivalent equation system of the geometric constraint system. The translation process is essentially the process separating the parameter $p$ from entities' coordinates $X$. Systematically, the geometric constraints are classified into four categories, i.e., dimensional constraint, structural constraint, algebraic constraint, and integral constraint. The first category includes constraints that are trivially separable; for the second and third categories, there is no need to carry out separation (but they do have different characteristics, thereby yielding two categories); the last category is not considered in the present work. The following provides detailed descriptions on the definitions and separability of these categories.

1) \textbf{Dimensional constraints}, includes the commonly used distance, angle, and size (e.g. radius/diameter). Their corresponding algebraic equations have been listed in Table \ref{tab:DimensionalConstraints} where $p$ denotes the parameter of its belonging constraint. In a general form, all the equations can be abstracted/rewritten as:
	\begin{align}
    \text{$p=f(X)$ $(p\geq 0)$,}
	\label{eq:4-1-1}
	\end{align}
which is already in a separation form, or
 	\begin{align}
    \text{$\cos{p}=f(X)$ $(0\leq p\leq \pi)$,}
	\label{eq:4-1-2}
	\end{align}
which can also be separated via a simple conversion:
 	\begin{align}
    \text{$p=\arccos{f(X)}$ $(0\leq p\leq \pi)$.}
	\label{eq:4-1-3}
	\end{align}
Basically, these equations have a very convenient form for separation: $p$ and $X$ reside at respective sides of the equation, and $p$ is only involved in one term. This special form makes the separation trivial, and Implicit Function Theorem is directly applicable.

\renewcommand{\arraystretch}{1.7} 
\begin{table}[width=.9\linewidth,cols=4,pos=h]
\caption{Dimensional constraints and their related equations.}
\label{tab:DimensionalConstraints}
\begin{tabular*}{\tblwidth}{@{} LLLL@{} }
 \hline
    Constraint types & Entities ($P$: point; $L$: line; $C$: circle) & Equations\\
    \hline
    Point-point distance & $P_1(x_1,y_1), P_2(x_2,y_2)$ & $p=\sqrt{(x_1-x_2)^2+(y_1-y_2)^2}$ \\
    Point-line distance & $P(x,y), L(a,b,c)$ &  $p=\sqrt{\frac{(ax+by+c)^2}{a^2+b^2}}$ \\
    Line-line distance & $L_1(a_1,b_1,c_1), L_2(a_2,b_2,c_2)$ & \begin{tabular}[c]{@{}l@{}} $p=\sqrt{\frac{(c_1-c_2)^2}{a_1^2+b_1^2}}$,  $a_1=a_2$, $b_1=b_2$\end{tabular}  \\
    Line-line angle & $L_1(a_1,b_1,c_1), L_2(a_2,b_2,c_2)$ & $\cos{p}=\frac{a_1a_2+b_1b_2}{\sqrt{a_1^2+b_1^2}\sqrt{a_2^2+b_2^2}}$ \\
    Circle radius & $C(x,y,r)$ & $p=r$ \\
    Circle diameter & $C(x,y,r)$ & $p=2r$\\
    \hline
\end{tabular*}
\end{table} 

2) \textbf{Structural constraints.} This category contains geometric constraints that have no parameters, e.g. perpendicular, coincidence, parallel, etc. Commonly used structural constraints are listed in Table \ref{tab:StructuralConstraints}. As there are no parameters involved, separation is not needed for them.

\renewcommand{\arraystretch}{1.7} 
\begin{table}[width=.9\linewidth,cols=4,pos=h]
\caption{Common structural constraints and their related equations.}
\label{tab:StructuralConstraints}
\begin{tabular*}{\tblwidth}{@{} LLLL@{} }
 \hline
    Constraint types & Entities ($P$: point; $L$: line; $C$: circle) & Equations\\
    \hline
    Point-point coincidence & $P_1(x_1,y_1), P_2(x_2,y_2)$ & $x_1=x_2$, $y_1=y_2$ \\
    Line-line coincidence & $L_1(a_1,b_1,c_1), L_2(a_2,b_2,c_2)$ & $a_1b_2=a_2b_1$,  $c_1a_2=c_2a_1$  \\
    Point-on-line & $P(x,y), L(a,b,c)$ &  $ax+by+c=0$ \\
    Point-on-circle & $P(x_1,y_1), C(x_2,y_2,r)$ &  $r=\sqrt{(x_1-x_2)^2+(y_1-y_2)^2}$ \\
    Line-line parallel & $L_1(a_1,b_1,c_1)$, $L_2(a_2,b_2,c_2)$ & $a_1b_2=a_2b_1$ \\
    Line-line perpendicular & $L_1(a_1,b_1,c_1), L_2(a_2,b_2,c_2)$ & $a_1a_2+b_1b_2=0$ \\
    Line-circle tangent & $L(a,b,c)$, $C(x,y,r)$ & $r=\sqrt{\frac{(ax+by+c)^2}{a^2+b^2}}$ \\
    Circle-circle tangent internally & $C_1(x_1,y_1,r_1)$, $C_2(x_2,y_2,r_2)$ & $(x_1-x_2)^2+(y_1-y_2)^2=(r_1-r_2)^2$\\
    Circle-circle tangent externally & $C_1(x_1,y_1,r_1)$, $C_2(x_2,y_2,r_2)$ & $(x_1-x_2)^2+(y_1-y_2)^2=(r_1+r_2)^2$ \\
    Vector normalization & $L(a,b,c)$ & $a^2+b^2=1$\\
    \hline
\end{tabular*}
\end{table} 

3) \textbf{Algebraic constraints.} This category includes constraints that are used to related different parameters using algebraic equations, e.g., $d_1^2+d_2^2=1$, where $d_1$ and $d_2$ are two distance parameters. In a general form, the algebraic constraint equations can be abstracted as:
	\begin{align}
    \text{$F_\text{A}(P)=0$,}
	\label{eq:4-1-4}
	\end{align}
where $P$ denotes the parameters. The parameters involved are defined by corresponding dimensional constraints beforehand; and these dimensional constraints are trivially separable, as already shown in their descriptions above. Their separated equations take the following general form:
	\begin{align}
    \text{$P=F(X)$.}
	\label{eq:4-1-5}
	\end{align}
By substituting this into Equation~(\ref{eq:4-1-4}), the latter can be converted to:
	\begin{align}
    \text{$F_\text{A}(F(X))=0$.}
	\label{eq:4-1-6}
	\end{align}
where there are no parameters involved, thus separation is not needed anymore.

4) \textbf{Integral constraints.} This category involves integral-based constraints such as constraints on area, solid volume, position of center of mass etc. Integral constraints cannot be handled by the proposed algorithm because, for it to work, all constraints must be convertible to algebraic equations and then separated. Unfortunately, integral constraints have no algebraic forms and cannot be separated in general. Hence, this category is not considered in the present work.

\begin{figure*}[t]
	\centering
    \includegraphics[scale=0.8]{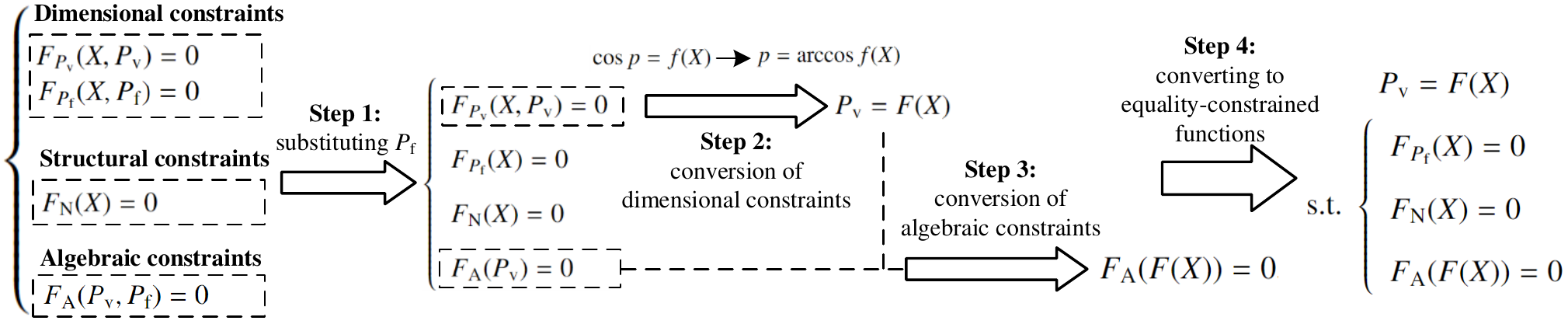}
	\caption{Illustration of the symbolic procedure.}
\label{fig:4-symbolic_procedure}
\end{figure*}

The symbolic procedure is then carried out based on the descriptions above. Basically, it is divided into four main steps:  (1) substitute the values of the fixed parameters into the equation system; (2) convert the dimensional constraint equations into the functional expressions of the variable parameters; (3) substitute the functional expressions of variable parameters into the equations of algebraic constraints; and (4) construct the equality-constrained functions of the variable parameters. An illustration of the procedure is presented in Figure~\ref{fig:4-symbolic_procedure}. In this figure, the parameters $P$ is divided into two groups: the variable parameters denoted as $P_\text{v}$, and the fixed parameters denoted as $P_\text{f}$.

\begin{figure*}[t]
	\centering
	\subfigure[]{\includegraphics[scale=0.8]{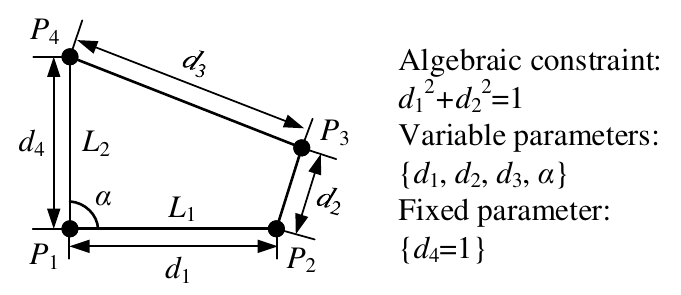}}\\
    \subfigure[]{\includegraphics[scale=0.7]{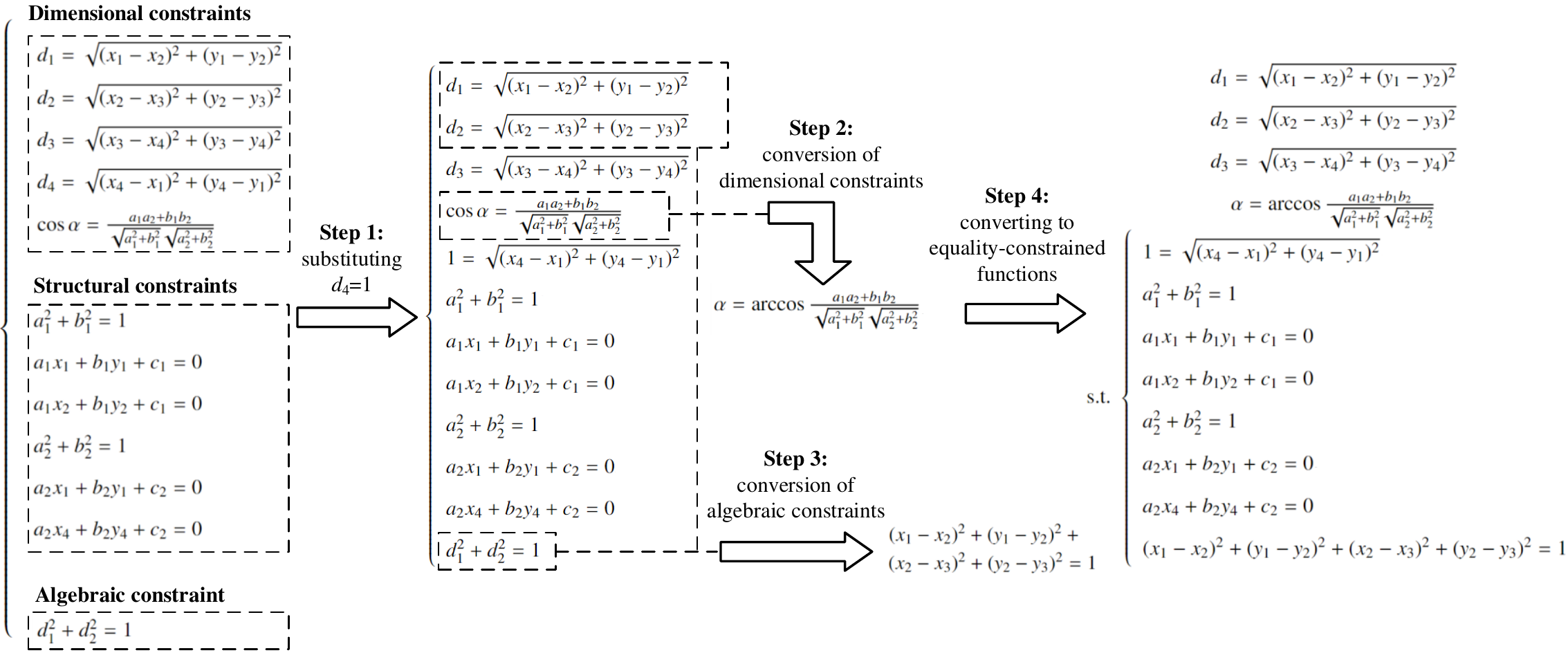} }
	\caption{An example of the symbolic procedure: (a) a geometric constraint system; and (b) the symbolic procedure to construct equality-constrained functions for the variable parameters.}
\label{fig:example of symbolic procedure}
\end{figure*}

A demonstrative example of the symbolic procedure is presented in Figure \ref{fig:example of symbolic procedure}. The geometric constraint system consists of 6 geometric entities $P_1(x_1,y_1)$, $P_2(x_2,y_2)$, $P_3(x_3,y_3)$, $P_4(x_4,y_4)$, $L_1(a_1,b_1,c_1)$, and $L_2(a_2,b_2,c_2)$; and 5 dimensional constraints $d_1$, $d_2$, $d_3$, $d_4$, and $\alpha$; and an algebraic constraint $d_1^2+d_2^2=1$; and several point-on-line structural constraints. Suppose $d_1$, $d_2$, $d_3$, and $\alpha$ are selected by the user as variable parameters, and $d_4$ stays unchanged as $1$. As the line segments between $P_2P_3$ and $P_3P_4$ can be directly determined by their end points, they are not defined here for brevity. Here, the one-dimensional ranges of the four variable parameters are to be computed. The symbolic procedure for deriving the equality-constrained functions of the four variable parameters is demonstrated in Figure~\ref{fig:example of symbolic procedure}(b).

\section{Calculation of closed endpoint candidates}

This section aims at calculating closed endpoint candidates of the function range intervals. First, the problem is translated into a local optimization problem; all local optimum values of the function are to be calculated as the values of closed endpoint candidates. Then, to determine all local optimums, Lagrange multipliers are introduced to construct an equation system, which is further solved by the NichePSO algorithm.

\subsection{Local optimization problem translation}

A closed endpoint of an interval, if it exists, corresponds to a local maximum or minimum of the equality-constrained function. A proof for the statement is presented as follows.

\textbf{Theorem.} Let $p$ be a parameter of a given geometric constraint system. Let $D$ be the range of $p$, and $p$ is expressed as
    \begin{equation}\nonumber
    \begin{array}{c}
    p=f(X),\\
    \text{ s.t. }  G(X)=0.
    \end{array}
    \end{equation}
where $f(X)$ and $G(X)$ are continuous functions. Then, any closed endpoints of $D$ are the conditional extremes of $f(X)$.

\textbf{Proof.} Might as well let $p_0$ be an upper closed endpoint of an interval in $D$. Let $X_0$ be the variable values satisfying
    \begin{equation}\nonumber
    \begin{array}{c}
    p_0=f(X_0),\\
    \text{ s.t. }  G(X_0)=0.
    \end{array}
    \end{equation}
Within the continuous neighborhood of $X_0$ in $G(X)=0$, any $X$ have $f(X)\leq p_0$. Thus, $p_0$ is a local maximum of $f(X)$. Similarly, the lower closed endpoints of $D$ are local minimums of $f(X)$. In summary, the closed endpoints of $D$ are conditional extremes of $f(X)$.

It is worthy of noting that a local optimum of the function does not necessarily correspond to a closed endpoint of the entire function range. The problem of calculating all closed endpoints can thus be converted to a local optimization problem where all local optimums are to be calculated as closed endpoint candidates.

\subsection{Determination of all local optimums}

Algorithm \ref{alg1:first} is presented to calculate all local optimums of the equality-constrained function. In this algorithm, first, the Lagrange multipliers are introduced to construct an equation system. Then, the equation system is solved by the NichePSO algorithm.

\begin{algorithm}
\caption{Determination of all local optimums} 
\label{alg1:first}
\begin{algorithmic}[1]
\REQUIRE ~~\\
$f$ - the equality-constrained function.
\ENSURE ~~\\
$V$ - the set of local optimum values.
\STATE $L\Leftarrow$ ConstructLagrangeFunction($f$) 
\STATE $E\Leftarrow$ ConstructEquationSystem($L$) 
\STATE $h\Leftarrow$ TransformToOptimizaitionFunction($E$) 
\STATE $R\Leftarrow$ NichePSO($h$) 
\FOR{each root $r \in R$}
\STATE $x\Leftarrow$ GetX($r$) 
\STATE $v\Leftarrow$ $f(x)$ 
\IF{NotExist($v,V$)}
\STATE add $v$ to $V$
\ENDIF
\ENDFOR
\RETURN $V$
\end{algorithmic}
\end{algorithm}

\subsubsection{Equation system construction with Lagrange multipliers}

The Lagrange multipliers method has been one of the most common methods for calculating local optimums of an equality-constrained function. In this paper, Lagrange multipliers are introduced to function constraints, and an equation system with the roots corresponding to the local optimums is constructed.

Regarding Equation~(\ref{eq:3-2}), let the size of $X$ be $m$, and let the number of equations in $G(X)=0$ be $n$. A Lagrange multiplier $\lambda_i$ is introduced to each constraint equation $g_i(X)=0$ from $G(X)=0$. Let $\Lambda$ be the set containing all Lagrange multipliers, the Lagrange function is then constructed as follows:
	\begin{align}
    L(X,\Lambda)=
    f(X)+
    \sum_{i=1}^n \lambda_ig_i(X).
	\end{align}
By letting all partial derivative functions of $L$ be zero, a system of nonlinear equations can be constructed as follows:
\begin{align}
	\left\{
	    \begin{array}{l}
    	\frac{\partial L}{\partial x_1}=0,\\
    	\cdots \\
    	\frac{\partial L}{\partial x_m}=0, \\
    	g_1(X)=0, \\
    	\cdots\\
    	g_n(X)=0.
    	\label{eq:5-3}
    	\end{array}
    \right.
\end{align}
The roots of Equation~(\ref{eq:5-3}) correspond to the local optimums of Equation~(\ref{eq:3-2}).

\subsubsection{Equation system solving}

The equation solving problem is converted into an optimization problem as follows:
	\begin{align}
    min. \quad h(X,\Lambda)=
    \sum_{i=1}^m (\frac{\partial L}{\partial x_i})^2 +
    \sum_{j=1}^n g_j^2.
    \label{eq:5-7}
	\end{align}
The NichePSO algorithm is then employed to solve this multi-modal function. 

The NichePSO algorithm is a variant of the classical PSO algorithm, a bio-inspired metaheuristic proposed by Kennedy and Eberhart \citep{kennedy1995pso}. In PSO, a population (called a swarm) of candidate solutions (called particles) is involved to perform a population-based search \citep{jafari2017fluid}. Each particle is characterized by position and velocity, which are randomly initialized within a limited range, and update according to its individual best-known position and swarm's best-known position. The update of velocity and position of particles are generally formulated as follows:
	\begin{align}
    \left\{
	    \begin{array}{ll}
    	v(t+1)=v(t)+c_1\phi_1(p(t)-x(t))+c_2\phi_2(g(t)-x(t)),\\
    	x(t+1)=x(t)+v(t+1),\\
    	\end{array}
    \right.
	\label{eq:5-8}
	\end{align}
where $v$ denotes particle velocity, $x$ denotes particle position, $t$ denotes the current number of iteration; $p$ is the individual best-known position, $g$ is the global best-known position, $c_1$, $c_2$ are adjustable parameters, $\phi_1$ and $\phi_2$ are randomly generated values in the interval $[0,1]$. The application of the update equation is repeated until a specified number of iterations has been reached, or until the updated velocity is close to zero.

NichePSO \citep{brits2002niching}, designed from the classical PSO, is one of the multi-modal optimization methods for finding multiple solutions to optimization problems \citep{lim2009innovations,goldberg1987genetic,goldberg1987genetic,esquivel2003use,kennedy2000stereotyping,li2002species}. It is a PSO-based algorithm combining with the niching technique. In this algorithm, a main swarm containing all particles are initialized; by tracking the movements of these particles, multiple subswarms (called niches) are identified and generated to search multiple solutions; during the searching process, subswarms will absorb the particles moving towards them, and will merge when intersecting with other subswarms. Several critical steps are clarified as follows:

1) \textbf{Initialization} The main swarm consisting of all particles is initialized for further partition. Since the success of the NichePSO relies on the proper initial distribution of particles in the search space, \textit{Faure}-sequences \citep{thiemard1998economic} is employed to guarantee the uniform distribution of the initial particles.

2) \textbf{Identification of niches} Niches are identified by keeping track of the changes in the fitness values of particles. If the individual best fitness of a particle does not change for a certain number of iterations, a niche is identified where there could be solutions around, and a new subswarm is generated containing the particle and its closed neighbors. 

3) \textbf{Merging subswarms} Two subswarms are merged if they are close in ($m$+$n$)-dimensional Euclidean space. The purpose of this step is to prevent different subswarms from converging to the same solutions.

In this paper, the NichePSO is suitable for resolving the problem for the following reasons:

(1) The number of solutions is difficult, if not impossible, to be determined in advance; the NichePSO, however, can adaptively generate sufficient subswarms without knowing the size of solutions beforehand.

(2) Special solutions, where the system is at a singular configuration, could be encountered. In the singular configuration, the Jacobian matrix of the equation system suffers rank reduction, resulting in a set of continuously distributed roots corresponding to the same value of the variable parameter. This situation can hardly be resolved by classical solving methods, e.g. homotopy continuation \citep{Allgower1993homotopy}, since they require discretely distributed roots. However, NichePSO algorithm is able to generate discrete solutions samples within a continuous solution set (only one solution is needed to compute the value of the variable parameter), thus can handle the singular situations. 

\begin{figure}[t]
    \centering
    \subfigure[]{\includegraphics[scale=0.8]{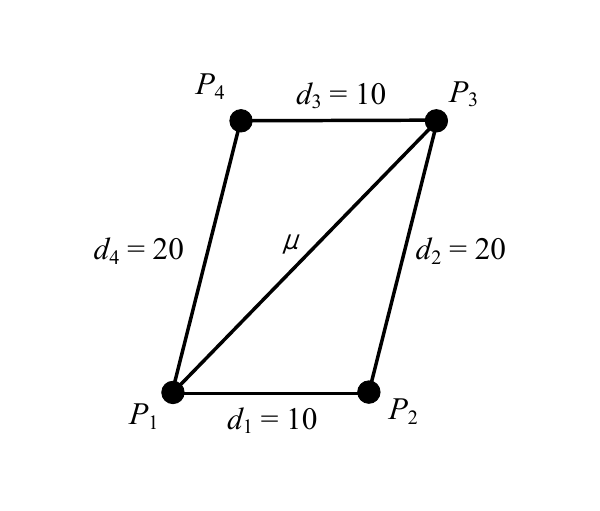}}
    \subfigure[]{\includegraphics[scale=0.8]{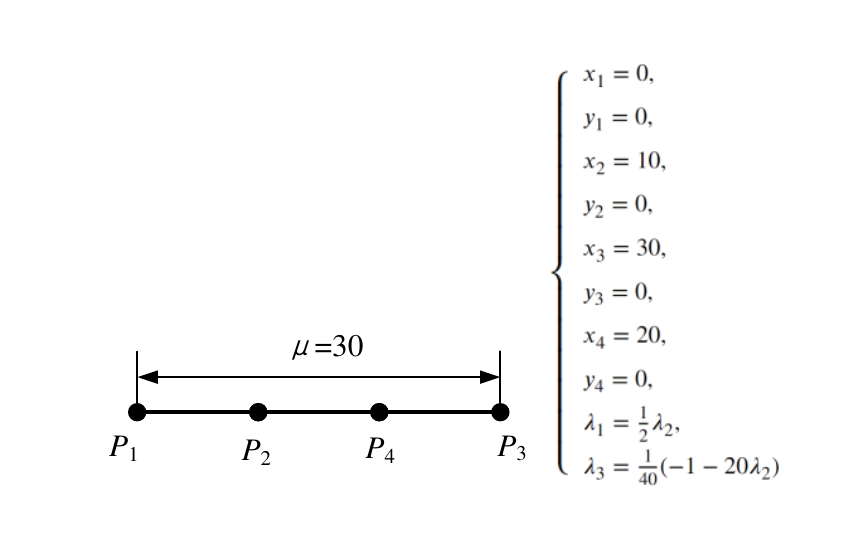}}
    \subfigure[]{\includegraphics[scale=0.8]{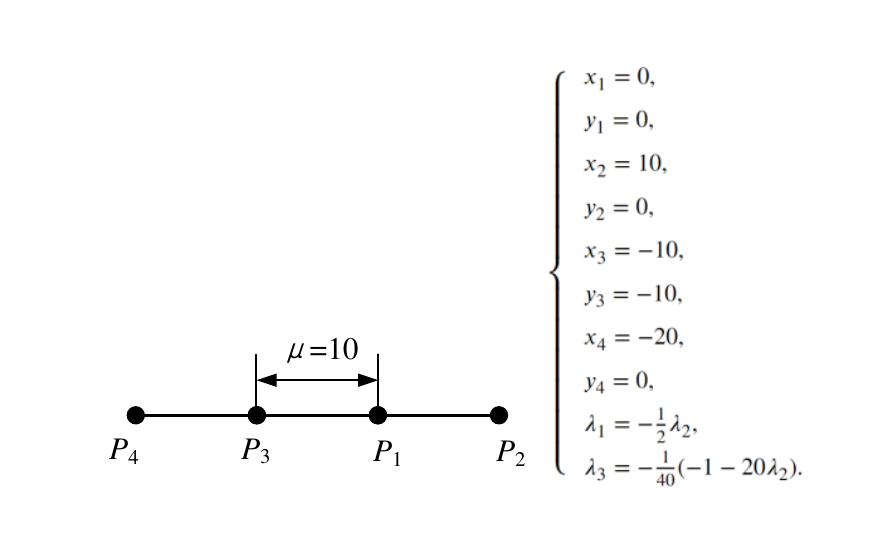}}
    \caption{Singular configuration at parameter optimums: (a) a geometric constraint system; (b) the singular configuration and the variable values when $\mu$ reaches maximum; and (c) the singular configuration and the variable values when $\mu$ reaches minimum.}
    \label{fig:5-3}
\end{figure}

An example is presented in Figure \ref{fig:5-3} for supporting the statement in reason (2). In the figure, $\mu$ denotes the only variable parameter to be edited. To calculate closed endpoint candidates of $\mu$, symbolic techniques are first adopted to solve the derived Lagrange equation system; the roots with the corresponding configuration are presented in Figure \ref{fig:5-3}(b-c) for reference. The results show that there are linear relationships among the Lagrange multipliers; thus the solutions are two continuously distributed sets, corresponding to the two optimum values of $\mu$ respectively ($10$ and $30$). Then, the NichePSO algorithm is used to solve the Lagrange equation system again. Finite number of solutions are obtained, and some of them are listed in Table \ref{tab:NichePSO}. The results show that the solutions are discretely distributed within the two continuous sets from Figure \ref{fig:5-3}(b-c). Based on these discrete solutions, the optimum values of $\mu$ can then be trivially carried out.

\renewcommand{\arraystretch}{1.5} 
\begin{table}[width=.6\linewidth,cols=4,pos=h]
  \caption{Solutions searched by the NichePSO algorithm.}
  \label{tab:NichePSO}
  \begin{tabular}{cccccccc}
    \hline
    Number & $x_3$ & $y_3$ & $x_4$ & $x_4$ & $\lambda_1$ & $\lambda_2$ &  $\lambda_3$\\
    \hline
    $1$ & $30$ & $0$ & $20$ & $0$ & $1.2$ & $0.6$ & $-0.325$\\
    $2$ & $30$ & $0$ & $20$ & $0$ & $-2.12$ & $-1.06$ & $0.555$\\
    $3$ & $-10$ & $-10$ & $-20$ & $0$ & $0.6$ & $-0.3$ & $-0.125$\\
    $4$ & $-10$ & $-10$ & $-20$ & $0$ & $1.61$ & $-0.805$ & $-0.4275$\\
    $5$ & $-10$ & $-10$ & $-20$ & $0$ & $3.82$ & $-1.96$ & $-0.955$\\
    $6$ & $-10$ & $-10$ & $-20$ & $0$ & $-4.36$ & $2.18$ & $1.115$\\
    $7$ & $-10$ & $-10$ & $-20$ & $0$ & $-0.04$ & $0.02$ & $0.035$\\
    $\cdots$ & $\cdots$ & $\cdots$ & $\cdots$ & $\cdots$ & $\cdots$ & $\cdots$ & $\cdots$ \\
    \hline
  \end{tabular}
\end{table}

\section{Calculation of open endpoint candidates}

The geometric constraint system could encounter singular cases where the two points defining a line coincide. The line in this situation is undefined and the geometric constraint system is thus unsolvable. Therefore, to guarantee the solvability of the geometric constraint system, the coordinate values of the geometric objects corresponding to the singular cases are taken as discontinuity points. If a discontinuity point happens to be at the modal of the equality-constrained function, the limit function value in its neighborhood could be the value of an open endpoint.

To calculate the open endpoint candidates, singular cases where the lines are undefined needs to be defined first. Given a line $L$ defined by two points $P_1(x_1,y_1)$ and $P_2(x_2,y_2)$, the singular case of the line can be expressed as follows:

	\begin{align}
	\begin{array}{c}
	(x_{1}-x_{2})^2+(y_{1}-y_{2})^2=0.
	\end{array}
	\label{eq:6-1}
	\end{align}
For the $k$ lines defined by points in a geometric constraint system, the singular case of the whole geometric constraint system where one of the lines is undefined can be expressed as follows:

	\begin{align}
	\begin{array}{c}
	C(X)=\prod_{i=0}^k c_i(X)=0,\\
	\text{where  } c_i(X)=(x_{i1}-x_{i2})^2+(y_{i1}-y_{i2})^2,
	\end{array}
	\label{eq:6-2}
	\end{align}
and $(x_{i1},y_{i1})$, $(x_{i2},y_{i2})$ are the coordinates of the two points defining the $i$th line.

The evaluation of equality-constrained function under the limit condition of singular cases can be expressed as follows:

	\begin{align}
	\begin{array}{c}
	p=\lim\limits_{C(X)\to0}f(X),\\
	\text{s.t. } G(X)=0.
	\end{array}
	\label{eq:6-3}
	\end{align}
By setting the value of $C(X)$ to a threshold value $\delta$, Equation~(\ref{eq:6-3}) can be transformed into:

	\begin{align}
	    \begin{array}{c}
                p=f(X),\\
                \text{s.t. } 
                \begin{cases}
                    G(X)=0, \\
                    C(X)=\delta.
                \end{cases}
        \end{array}
        \label{eq:6-4}
	\end{align}
The evaluation of the function in Equation~(\ref{eq:6-4}) could be finite values, closed intervals, or even an empty set. The evaluation is an empty set when there is conflict between the new-added constraint $C(X)=\delta$ and the original constraint equations $G(X)=0$; the evaluation is a finite set of values when the state of the equation set composed of the equality constraints is well-constrained; and the result is a set of closed intervals when the state of the equation set composed of equality constraints is under-constrained. Regarding the case where the evaluation is a set of closed intervals, only their endpoints need to be calculated as the open endpoint candidates. The calculation of these endpoints can be achieved by the method in in Section 5.

\section{Experimental results and limitations}

The proposed approach has been verified by experiments. The algorithm presented in this paper has been implemented using MATLAB 2019b software on a computer with an Intel Core i5-8500 CPU at 3.0 GHz and 16-GB RAM, and was tested on different 2D geometric constraint systems. Several representative case studies were conducted to evaluate the effectiveness and efficiency of the proposed approach.

\subsection{Case studies}

\setlength{\parskip}{0.5\baselineskip}\noindent\textit{Case 1: A ruler-and-compass constructible geometric constraint system}

\begin{figure*}[t]
	\centering
	\subfigure[]{\includegraphics[scale=0.8]{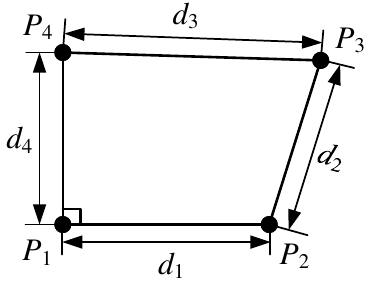}}  
	\hspace{35pt}
    \subfigure[]{\includegraphics[scale=0.8]{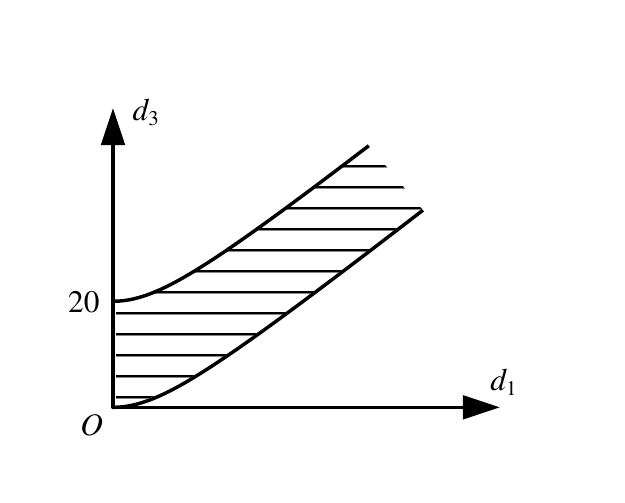}} \\
    \subfigure[]{\includegraphics[scale=0.8]{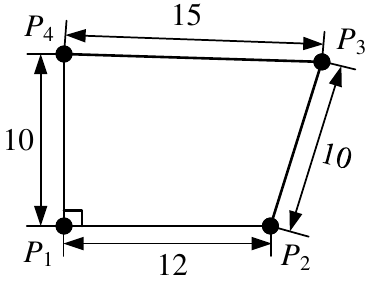}}
    \hspace{35pt}
    \subfigure[]{\includegraphics[scale=0.8]{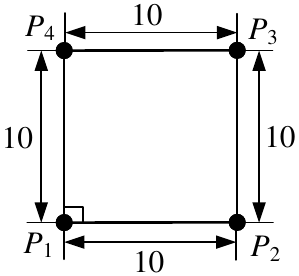}}  
	\hspace{35pt}
    \subfigure[]{\includegraphics[scale=0.8]{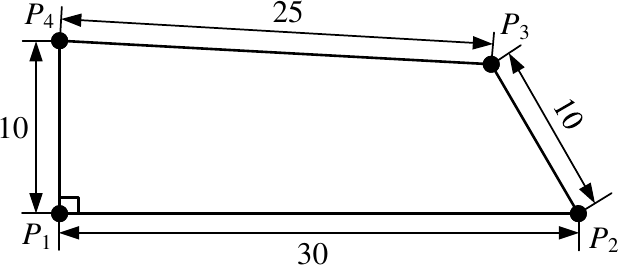}} 
	\caption{Illustration of multi-parameter editing in a ruler-and-compass constructible geometric constraint system: (a) the geometric constraint system; (b) the parameter space related to $d_1$ and $d_3$ (dashed region); (c) the original model; (d) edited model by the constructive method \citep{van2005constructive} and the first editing operation of the proposed approach; and (e) edited model by the second editing operation of the proposed approach.}
    \label{fig:8-1}
\end{figure*}

The first case is presented to evaluate the approach's effectiveness of computing complete parameter ranges in multi-parameter editing. The geometric constraint system given in Figure \ref{fig:8-1}(a) is used to construct a 2D quadrangle. It consists of 4 points $P_1$, $P_2$, $P_3$, and $P_4$. The geometric constraints include 4 point-point distance constraints and 1 perpendicular constraint. The geometric constraint system can be decomposed into 2D triangular subproblems (triangle $P_1P_2P_4$ and triangle $P_2P_3P_4$), and thus is ruler-and-compass constructible. The initial values of parameters are as follows: $d_1=12$, $d_2=10$, $d_3=15$, and $d_4=10$; the original model is shown in Figure \ref{fig:8-1}(c); $d_1$ and $d_3$ are selected as the two variable parameters. The relation between $d_1$ and $d_3$ is trivially deduced as follows. First, in triangle $P_1P_2P_4$, the distance between $P_2$ and $P_4$ can be expressed as
	\begin{align}
    d_{P_2P_4}=\sqrt{d_1^2+d_4^2}.
    \label{eq:7-1}
	\end{align}
According to triangle inequalities, the parameter space related to all the parameters can be derived as:
	\begin{align}
    \text{$S_{P}=\{P:$}\left\{
             \begin{array}{ll}
             d_3 + d_2\geq \sqrt{d_1^2+d_4^2},  \\
             d_2 + \sqrt{d_1^2+d_4^2}\geq d_3,  \\
             \sqrt{d_1^2+d_4^2} + d_3\geq d_2,  \\
             d_1\geq0, \\
             d_2\geq0, \\
             d_3\geq0, \\
             d_4\geq0, \\
             \end{array}
    \right.\text{\}.}
    \label{eq:7-2}
	\end{align}
where $P=(d_1,d_2,d_3,d_4)$, and $S_P$ denotes the space of parameters. Substituting the values of $d_2$ and $d_4$, Equation~(\ref{eq:7-2}) is converted into:
	\begin{align}
    \text{$S_{P}'=\{P':$}\left\{
	    \begin{array}{l}
    	d_1^2-(d_3+10)^2+100 \leq 0,\\
    	d_1^2-(d_3-10)^2+100 \geq 0,\\
    	d_1\geq0, \\
        d_3\geq0, \\
    	\end{array}
    \right.\text{\}.}
    \label{eq:7-3}
	\end{align}
where $P'=(d_1,d_3)$, and $S_{P}'$ denotes the parameter space related to $d_1$ and $d_3$. Equation~(\ref{eq:7-3}) is depicted in Figure \ref{fig:8-1}(b) for reference.

\renewcommand{\arraystretch}{1.5} 
\begin{table}[]
  \centering
  \fontsize{6.5}{8}\selectfont
  \begin{threeparttable}
  \caption{The results in Case 1.}
  \label{tab:case1}
  \begin{tabular}{cccccc}
    \hline
    \multirow{2}{*}{Approach}  & \multirow{2}{*}{Parameter} & \multicolumn{2}{c}{1} & \multicolumn{2}{c}{2} \\ 
    \cmidrule(r){3-4} \cmidrule(r){5-6} 
    & & Range & Assignment & Range & Assignment \\
    \hline
    \multirow{2}{*}{\begin{tabular}[c]{@{}c@{}}Constructive approach  \end{tabular}}& $d_1$ & $[0,22.91]$ & $10$ & & \\
    & $d_3$ & $[5.62,25.62]$ &  & $[4.14,24.14]$ & $10$ \\
    \multirow{2}{*}{Our approach 1}  & $d_1$ & $[0,+\infty)$ & $10$ &  &  \\
    & $d_3$ & $[0,+\infty)$ &  & $[4.14,24.14]$ & $10$ \\
    \multirow{2}{*}{Our approach 2} & $d_1$ & $[0,+\infty)$ & $30$ & &  \\
    & $d_3$ & $[0,+\infty)$ & & $[21.62,41.62]$ & $25$ \\
    \hline
  \end{tabular}
  \end{threeparttable}
\end{table}

To conduct the comparison between the proposed approach and the constructive method by van der Meiden and Bronsvoort \citep{van2005constructive}, $d_1$ and $d_3$ are sequentially edited with the one-dimensional allowable ranges computed by the both approaches. The computed one-dimensional parameter ranges by the two approaches are given in Table \ref{tab:case1}, where the proposed approach is executed twice with different parameter assignments; the result models after editing are illustrated in Figure \ref{fig:8-1}(d-e).

\newpage

\setlength{\parskip}{0.5\baselineskip}\noindent\textit{Case 2: A geometric constraint system for constructing a hexagon}

\begin{figure*}[t]
    \centering
	\subfigure[]{\includegraphics[scale=0.8]{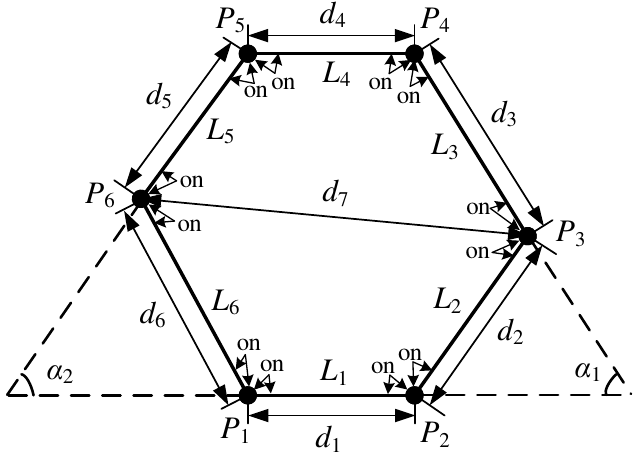}}  
    \subfigure[]{\includegraphics[scale=0.8]{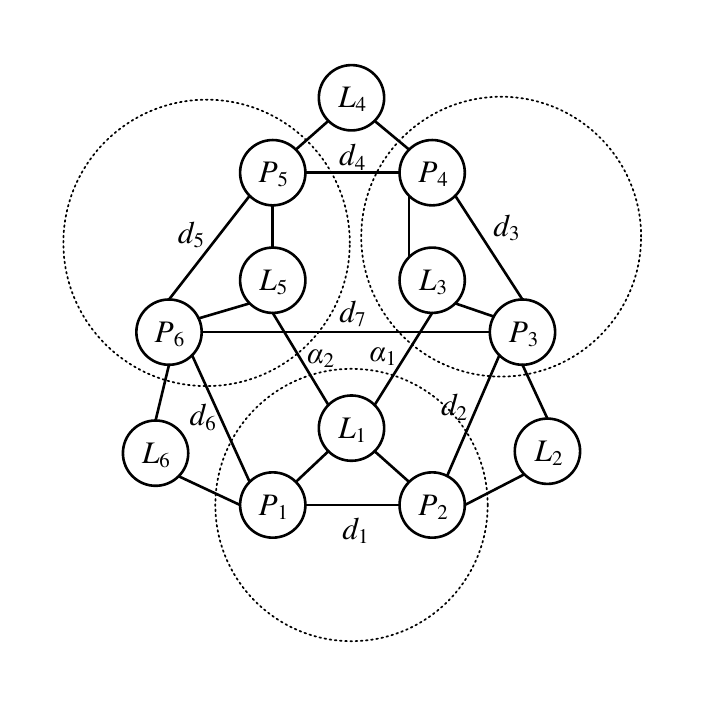}} 
    \subfigure[]{\includegraphics[scale=0.8]{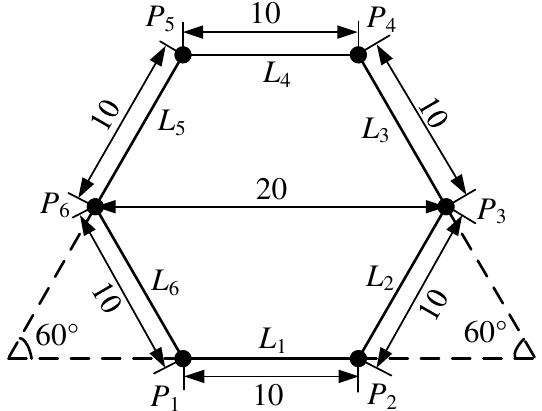}}
    \caption{Illustration of multi-parameter editing in a geometric constraint system that cannot be decomposed into triangular subproblems: (a) the geometric constraint system; (b) the graph representation of the geometric constraint system; and (c) result model by the proposed approach.}
\label{fig:8-2}
\end{figure*}

To show the effectiveness of our approach in handling general geometric constraint systems, the second case is tested, shown in Figure \ref{fig:8-2}. It is a geometric constraint system for constructing a 2D hexagon, consisting of 12 geometric entities, including 6 points and 6 lines, 9 dimensional constraints, including 7 distances and 2 angles, and 12 point-on-line structural constraints. The graph representation of the system is given in \ref{fig:8-2}(b). The system can be abstracted as a merge pattern of three rigid bodies, thus it can hardly be decomposed into triangular subproblems, and is not ruler-and-compass constructible. 

\renewcommand{\arraystretch}{1.5} 
\begin{table*}[]
  \centering
  \fontsize{6.5}{7}\selectfont
  \begin{threeparttable}
  \caption{Sequentially editing all parameters in the geometric constraint system from Case 2.}
  \label{tab:case2}
  \begin{tabular}{ccccccccccc}
    \hline
    \multirow{2}{*}{Parameter} & \multicolumn{2}{c}{1} & \multicolumn{2}{c}{2} & \multicolumn{2}{c}{3} & \multicolumn{2}{c}{4} & \multicolumn{2}{c}{5} \\ 
    \cmidrule(r){2-11}
    & Range & Assignment & Range & Assignment &  Range & Assignment &  Range & Assignment &  Range & Assignment  \\
    \hline
    $d_1$ & $[0,+\infty)$ & $10$ & \\
    $d_2$ & $[0,+\infty)$ & & $[0,+\infty)$ & $10$ & \\
    $d_3$ & $[0,+\infty)$ & & $[0,+\infty)$ & & $[0,+\infty)$ & $10$ &  \\
    $d_4$ & $[0,+\infty)$ & & $[0,+\infty)$ & & $[0,+\infty)$ & & $[0,+\infty)$ & & $[0,+\infty)$ & $10$ \\
    $d_5$ & $[0,+\infty)$ & & $[0,+\infty)$ & & $[0,+\infty)$ & & $[0,+\infty)$ & & $[0,+\infty)$ &  \\
    $d_6$ & $[0,+\infty)$ & & $[0,+\infty)$ & & $[0,+\infty)$ & & $[0,+\infty)$ & $10$  \\
    $d_7$ & $[0,+\infty)$ & & $[0,+\infty)$ & & $[0,+\infty)$ & & $[0,+\infty)$ & & $[0,30]$ & \\
    $\alpha_1$ & $[0,\pi)$ & & $[0,\pi)$ & & $[0,\pi)$ & & $[0,\pi)$ & & $[0,\pi)$ & \\
    $\alpha_2$ & $[0,\pi)$ & & $[0,\pi)$ & & $[0,\pi)$ & & $[0,\pi)$ & & $[0,\pi)$ & \\
    \hline
    \multirow{2}{*}{Parameter} &
    \multicolumn{2}{c}{6} & \multicolumn{2}{c}{7} & \multicolumn{2}{c}{8} & \multicolumn{2}{c}{9} \\
    \cmidrule(r){2-9}
    &  Range & Assignment &  Range & Assignment &  Range & Assignment &  Range & Assignment\\
    \cmidrule(r){1-9}
    $d_1$ \\
    $d_2$\\
    $d_3$\\
    $d_4$\\
    $d_5$ & $[0,50]$ & & $[0,40]$ & & $[0,38.3]$ & & $[0,33.7]$ & $10$\\
    $d_6$\\
    $d_7$ & $[0,30]$ & & $[0,30]$ & $20$ & &\\
    $\alpha_1$ & $[0,\pi)$ & $60^{\circ}$\\
    $\alpha_2$ & $[0,\pi)$ & & $[0,\pi)$ & & $[0,158.97)$& $60^{\circ}$ \\
    \cmidrule(r){1-9}
  \end{tabular}
  \end{threeparttable}
\end{table*}

In this case, all the parameters are selected as variable parameters and they are edited sequentially within the one-dimensional parameter ranges computed by the proposed approach. The computed one-dimensional parameter ranges are given in Table  \ref{tab:case2} and the result models after editing are illustrated in Figure \ref{fig:8-2}(c).

\setlength{\parskip}{0.5\baselineskip}\noindent\textit{Case 3: A geometric constraint system with conics}

\begin{figure*}[t]
    \centering
	\subfigure[]{\includegraphics[width=0.35\textwidth]{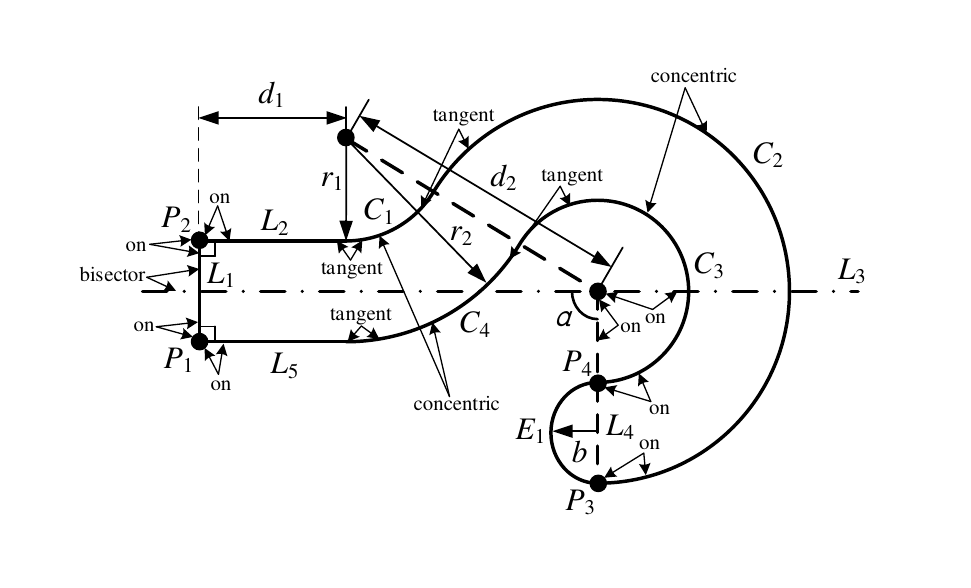}} 
    \subfigure[]{\includegraphics[width=0.32\textwidth]{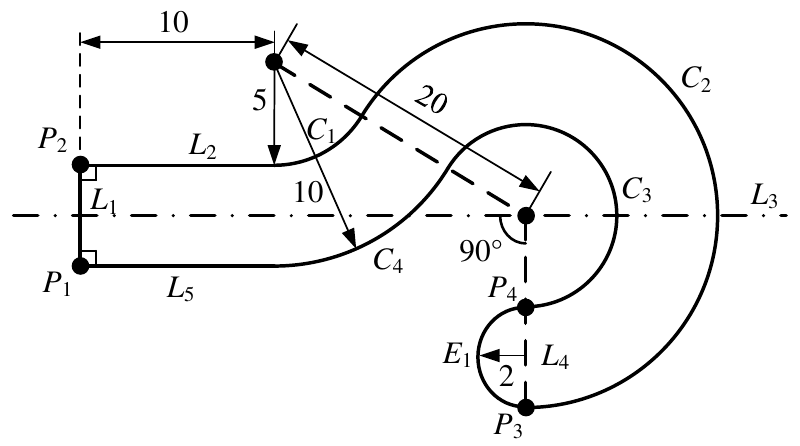}} 
    \subfigure[]{\includegraphics[width=0.3\textwidth]{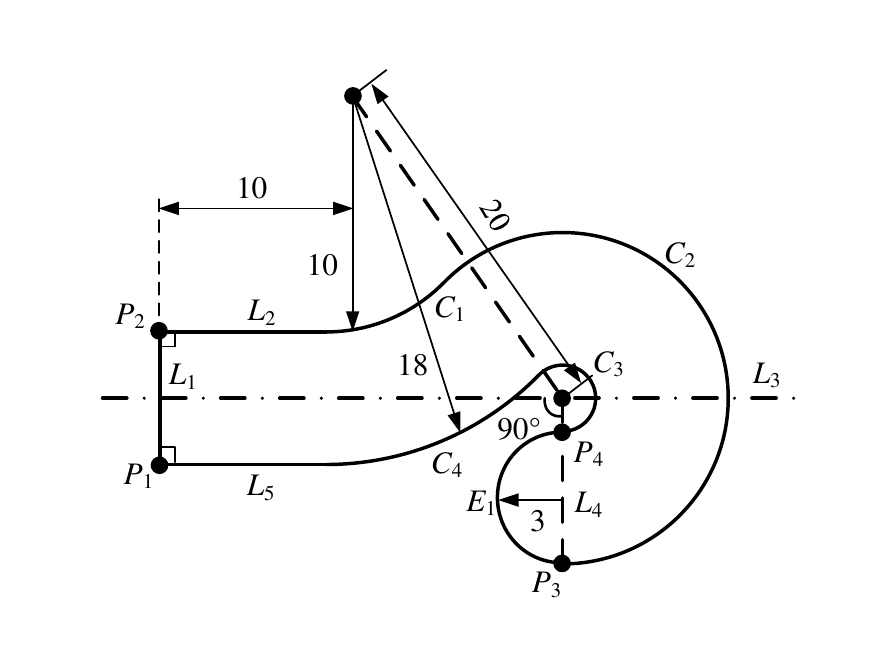}}
	\caption{Illustration of multi-parameter editing in a geometric constraint system with conics: (a) the geometric constraint system; (b) the original model; and (c) result model by the proposed approach.}
\label{fig:8-3}
\end{figure*}

To verify the effectiveness of the proposed approach in handling geometric constraint systems with conics, Case 3 is (presented in Figure \ref{fig:8-3}) was tested, where several circular arcs and an elliptic arc were included. Figure \ref{fig:8-3}(a) shows a geometric constraint system consisting of 15 geometric entities, including 4 points, 5 lines, 4 circular arcs, and 1 elliptic arc; and 6 dimensional constraints, including 2 distances, 1 angle, 2 radius constraints of the circular arcs, and 1 short radius constraint of the elliptic arc; and 21 structural constraints, including 8 point-on-line constraints, 2 point-on-circle constraints, 2 concentric constraints, 4 tangent constraints, 1 symmetric constraint, and 2 perpendicular constraints. The original model is shown in Figure \ref{fig:8-3}(b); parameters $r_1$, $r_2$, and $b$ are selected as the variable parameters and are edited in sequence. The computed parameter ranges are given in Table \ref{tab:case3} and the result model is illustrated in Figure \ref{fig:8-3}(c).

\renewcommand{\arraystretch}{1.5} 
\begin{table}[]
  \centering
  \fontsize{6.5}{8}\selectfont
  \begin{threeparttable}
  \caption{Sequentially editing $r_1$, $r_2$ and $b$ in the geometric constraint system from Case 3.}
  \label{tab:case3}
  \begin{tabular}{ccccccc}
    \hline
    \multirow{2}{*}{Parameter} & \multicolumn{2}{c}{1} & \multicolumn{2}{c}{2} & \multicolumn{2}{c}{3}\\
    \cmidrule(r){2-7}
    & Range & Assignment & Range & Assignment & Range & Assignment\\
    \hline
    $r_1$ & $[0,20]$ & $10$ & & & \\
    $r_2$ & $[0,20]$ & & $[0,20]$ & $18$ & &  \\
    $b$ & $[0,+\infty)$ & & $[0,+\infty)$ & & $[0,+\infty)$ & $3$  \\
    \hline
  \end{tabular}
  \end{threeparttable}
\end{table}

\setlength{\parskip}{0.5\baselineskip}\noindent\textit{Case 4: A large-scale geometric constraint system}

\begin{figure*}[t]
    \centering
    \subfigure[]{\includegraphics[width=0.4\textwidth]{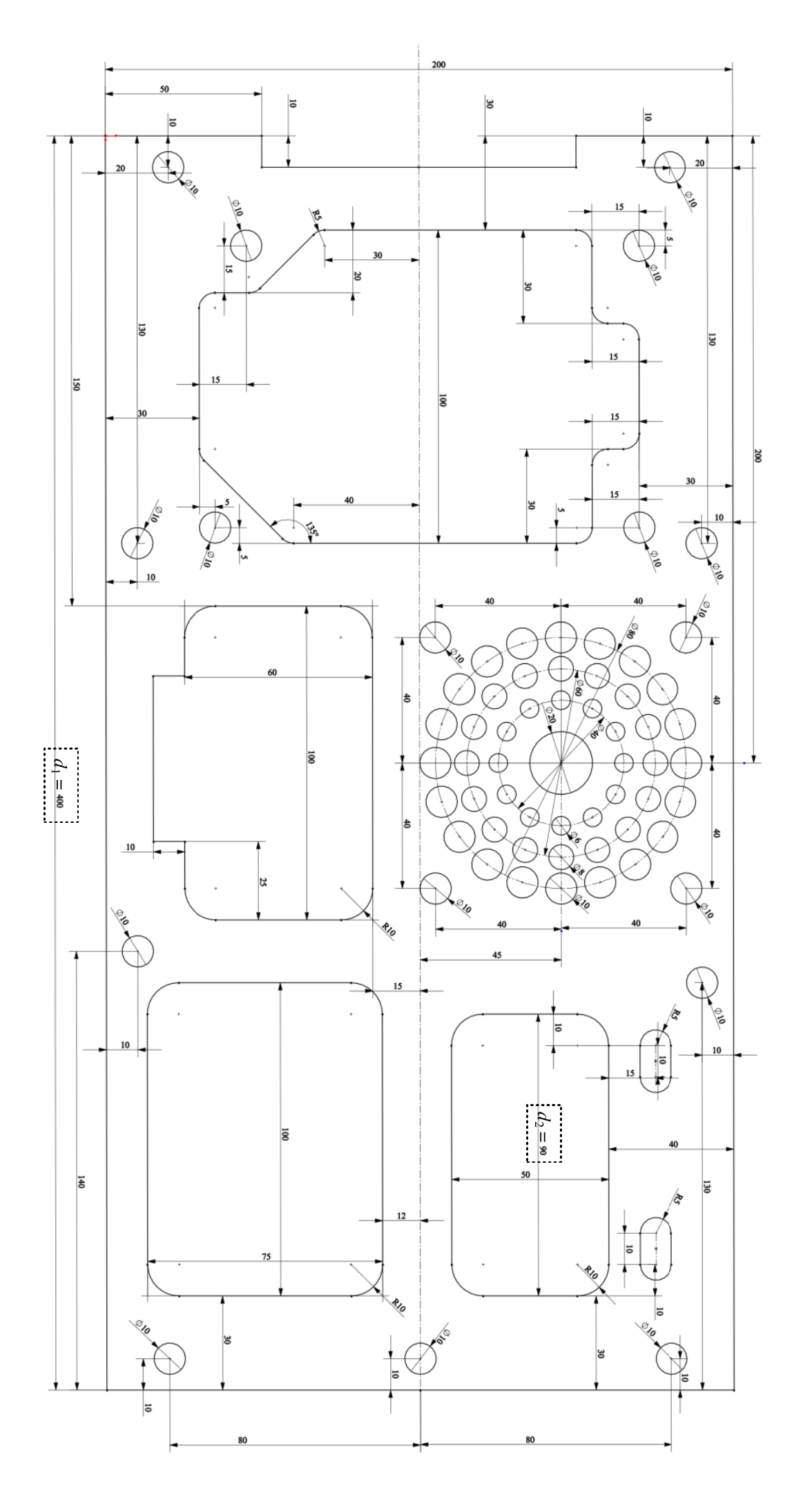}}
    \hspace{.5in}
    \subfigure[]{\includegraphics[width=0.4\textwidth]{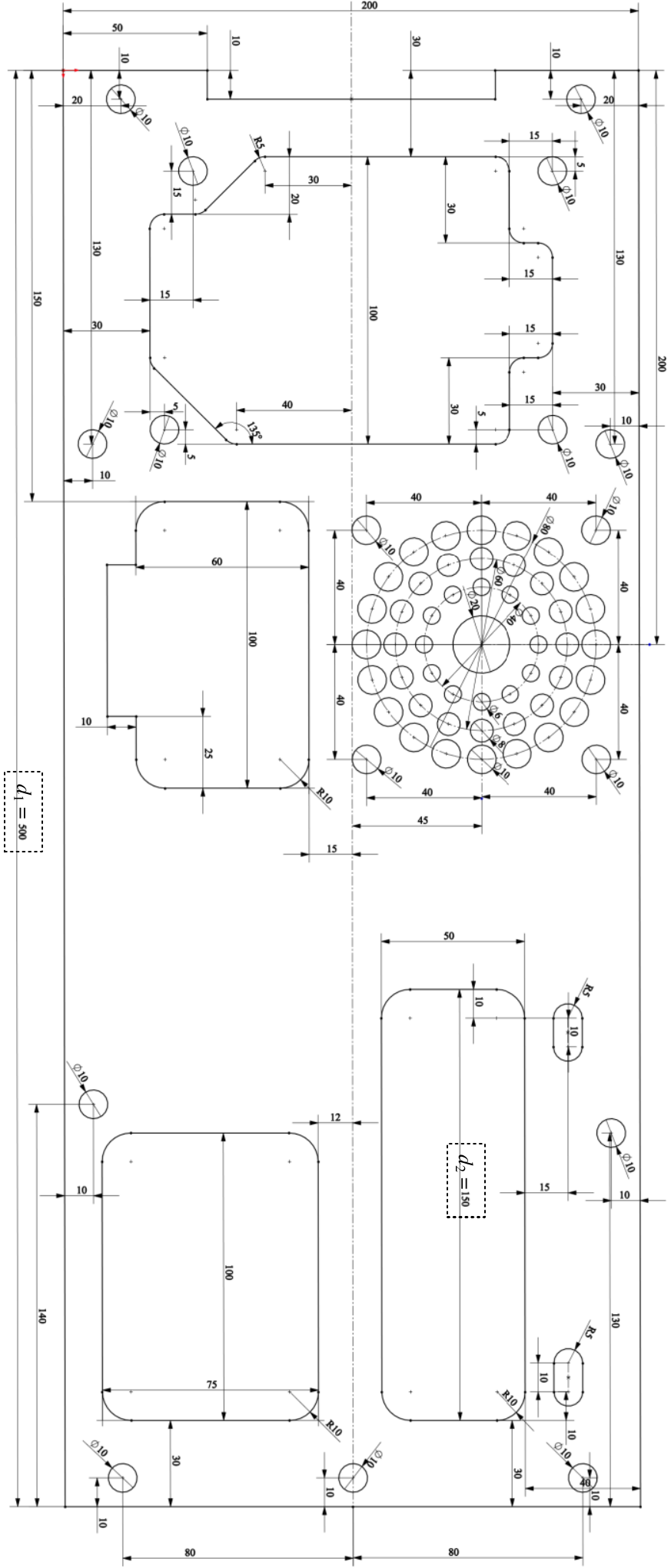}}
	\caption{Illustration of multi-parameter editing in a large-scale geometric constraint system: (a) the original model; (b) result model by the proposed approach.}
\label{fig:8-4}
\end{figure*}

To show the effectiveness of the proposed approach in handling large-scale problems, Case 4 is tested and presented in Figure \ref{fig:8-4}. It is a large-scale geometric constraint system consisting of 74 points, 96 circles or circular arcs, and 39 lines or line segments. Two of the parameters $d_1$ and $d_2$ (bounded by dashed boxes in Figure \ref{fig:8-4}(a)) are selected as variable parameters and edited sequentially with the one-dimensional parameter ranges calculated by the proposed approach. The computed parameter ranges are given in Table \ref{tab:case4}, and the result model is illustrated in Figure \ref{fig:8-4}(b). 

Moreover, to test the efficiency of the proposed approach, single-parameter editing is performed in each of the four cases, and the time spent is given in Table \ref{tab:time}. $5000$ particles are initialized in the main swarm, and the maximum number of iterations is set at $500$.

\renewcommand{\arraystretch}{1.5} 
\begin{table*}[]
  \centering
  \fontsize{6.5}{8}\selectfont
  \begin{threeparttable}
  \caption{Multi-parameter editing in the geometric constraint system from Case 4.}
  \label{tab:case4}
  \begin{tabular}{ccccc}
    \hline
    \multirow{2}{*}{Parameter} & \multicolumn{2}{c}{1} & \multicolumn{2}{c}{2}\\
    \cmidrule(r){2-5}
    & Range & Assignment & Range & Assignment \\
    \hline
    $d_1$ & $[0,+\infty)$ & $500$ &  \\
    $d_2$ & $[0,+\infty)$ & & $[0,+\infty)$ & $150$  \\
    \hline
  \end{tabular}
  \end{threeparttable}
\end{table*}

\renewcommand{\arraystretch}{1.5} 
\begin{table*}[tp]
  \centering
  \fontsize{6.5}{8}\selectfont
  \begin{threeparttable}
  \caption{Efficiency performance of the proposed approach.}
  \label{tab:time}
  \begin{tabular}{cccccccccc}
    \hline
    \multirow{2}{*}{Case} & \multirow{2}{*}{Variables} & \multirow{2}{*}{\begin{tabular}[c]{@{}c@{}}Total\\time\end{tabular}} & \multicolumn{3}{c}{Calculating closed endpoint candidates} & \multicolumn{3}{c}{Calculating open endpoint candidates} & \multicolumn{1}{c}{\begin{tabular}[c]{@{}c@{}}Determining\\ valid intervals\end{tabular}}\\
    \cmidrule(r){4-10}
    & & & Generated subswarms & Max iterations & Time & Generated subswarms & Max iterations & Time & Time \\
    \hline
    $1$ & $14$ & $24.89$ & $152$ & $500$ & $12.06$ & $136$ & $500$ & $11.66$ & $1.17$\\
    $2$ & $30$ & $41.95$ & $306$ & $500$ & $21.30$ & $288$ & $500$ & $19.21$ & $1.44$\\
    $3$ & $38$ & $47.79$ & $335$ & $500$ & $23.61$ & $316$ & $500$ & $22.67$ & $1.51$\\
    $4$ & $553$ & $570.65$ & $867$ & $500$ & $291.01$ & $732$ & $500$ & $276.01$ & $3.63$\\
    \hline
  \end{tabular}
  \end{threeparttable}
\end{table*}

\subsection{Discussions and limitations}

According to the results of Case 1 listed in Table \ref{tab:case1}, the instance of the geometric constraint system in Figure \ref{fig:8-1}(d) can be obtained by assigning parameter values within the allowable ranges computed by both approaches. The parameter assignments in the instance of the geometric constraint system in Figure~\ref{fig:8-1}(e) are within the allowable ranges computed by the proposed approach, but outside the allowable ranges computed by the constructive method \citep{van2005constructive}. This indicates that the parameter ranges computed by the constructive method \citep{van2005constructive} are incomplete. In addition, the allowable ranges computed by the proposed approach are consistent with the projection from the semi-algebraic set in Figure \ref{fig:8-1}(b) to corresponding axis, which means that the computed ranges are complete.

In terms of Case 2 and Case 3, the results show that the proposed approach can effectively handle the multi-parameter editing in these two general geometric constraint systems. As mentioned in \citep{sitharam2011cayley1,sitharam2011cayley2,sitharam2014beast}, the methods proposed by Sitharam et al. are restricted in geometric constraint systems with only point objects and distance parameters; and as pointed out in \citep{van2005constructive, hidalgo2012computing}, the geometric constraint systems considered in the constructive method are those with 2D triangular and 3D tetrahedral subproblems, points and straight lines as geometric objects, and point–point distances and line–line angles as geometric constraints. Although the considered geometric objects has been enlarged in \citep{hidalgo2012computing}, a systematic approach hasn't been proposed yet to handle geometric constraint systems with conics and radius constraints. Thus, based on our understanding, the existing methods can hardly deal with the parameter range problem in these two general cases.

From the results of Case 4, it can be concluded that the proposed approach can effectively handle large-scale problems but has efficiency issues at current stage. From the table, the bottle-neck of the algorithm’s timings is seen to be the NichePSO algorithm. The reason behind is that too many subswarms are generated during the searching phase to search only for several connected components within the solution space; and the particles remaining in the main swarm are difficult to converge. To address this, parameter tuning or heuristics can be used to merge the subswarms generated nearby the same connected component and to accelerate the convergence of the particles in the main swarm. There are several steps of the proposed method that can be speeded up by parallelization and cloud computing. Specifically, the searching of the particles initialized in the main swarm can be executed simultaneously; the searching of different subswarms can also be made parallel; in addition, the validation of each interval candidate can be done concurrently. 

The given cases are restricted to 2D geometric constraint systems. The extension of the proposed approach to 3D parametric modeling is considered as a more difficult issue to address. The primary challenge of this extension is that 3D parametric modeling is mainly procedural, which can hardly be converted to valid algebraic equations. Therefore, an equation-based representation scheme must be developed for 3D parametric modeling, and then the proposed method can be directly applicable. Moreover, inherent robustness issues (e.g. persistent naming \citep{farjana2016implementation, safdar2020feature}) in procedural modeling are also critical challenges to be faced with for 3D extension. challenge is the high computation load when applying the proposed method to 3D cases. To mitigate this issue, GPU parallel computing can be leveraged.

Moreover, the root selection problem \citep{bettig2003solution,kale2008geometric} has not been considered in the approach. The proposed approach, in its current form, can only guarantee the existence of roots, without ensuring that the desired root is contained in the computed parameter ranges. In general, the root selection is not involved in the numerical solving phase of geometric constraint solving, thus it is difficult to directly express them as an equation system. Therefore, the main difficulty to compute parameter ranges with the desired root lies in developing an equation-based representation scheme involving both the root selection rules and geometric constraints, based on which the proposed approach is applicable.

\section{Conclusions and future work}

Parametric modeling is widely applied in CAD software packages. However, current CAD systems are still unable to provide guidance of allowable parameter ranges for multi-parameter editing. A systematic approach has thus been presented in this paper to provide users with complete one-dimensional ranges of variable parameters in multi-parameter editing. In this approach, each variable parameter is represented as an equality-constrained function. The completeness of parameter ranges is guaranteed by maintaining a complete parameter space for projection, which is achieved by involving all the geometric constraints as constraint equations except for those related to unassigned variable parameters. The allowable range of each parameter to be edited is determined by calculating the range of the equality-constrained function where optimization methods are used. The order of editing parameters (up to the user) could affect the output parameter ranges. This is because editing different parameters makes the current space of variable parameters restricted into different dimensions, so it could affect the allowable ranges of the subsequent-edited parameters.

However, although the proposed approach can achieve good performances in multi-parameter editing, it still has the following issues to be addressed in future work.

\begin{enumerate}
    \item Efficiency. GPU parallel computing can be used to improve the efficiency of the proposed approach. Cloud computing techniques and cloud-based CAD techniques are also of great interests for future studies.
    
    \item 3D extension. To address this, the equation-based scheme of 3D parametric models needs to be developed first.
    
    \item Root selection. For this, a new equation-based scheme of geometric constraint solving is needed to combine the numerical solving phase together with the root selection phase.
\end{enumerate}


\bibliographystyle{model1-num-names}

\bibliography{cas-refs}


\end{document}